\documentclass[a4paper,11pt]{article}
\usepackage[skins]{tcolorbox}
\usepackage{mathtools,slashed}
\usepackage[T1]{fontenc}
\usepackage{slashed,verbatim,subfigure}
\usepackage[numbers,sort&compress]{natbib}
\usepackage{amsmath}
\usepackage{mathrsfs}
\usepackage{amsbsy}
\usepackage{amssymb}
\usepackage{appendix}
\usepackage{graphicx}
\usepackage[final]{pdfpages}
\usepackage{dcolumn}
\usepackage{bm}
\usepackage[colorlinks=true,linktocpage=true,
linkcolor=blue,citecolor=blue]{hyperref}
\usepackage[a4paper]{geometry}
\catcode`@11
\def\seceqaa{\@addtoreset{equation}{section}
	\def\theequation{A\arabic{equation}}}
\def\seceqbb{\@addtoreset{equation}{section}
	\def\theequation{B\arabic{equation}}}
\def\seceqcc{\@addtoreset{equation}{section}
	\def\theequation{C\arabic{equation}}}
\def\seceqdd{\@addtoreset{equation}{section}
	\def\theequation{D\arabic{equation}}}
\def\seceqee{\@addtoreset{equation}{section}
	\def\theequation{E\arabic{equation}}}
\catcode`@11
\labelformat{section}{Section #1} 
\labelformat{subsection}{Subsection #1} 
\labelformat{subsubsection}{Section #1}
\labelformat{subsubsubsection}{Section #1}
\labelformat{equation}{Eq.~(#1)} 
\labelformat{figure}{Fig.~#1} 
\labelformat{subfigure}{Fig.~\thefigure#1} 
\labelformat{table}{Tab.~#1} 
\labelformat{appendix}{Appendix #1}

\newcommand{\be}{\begin{eqnarray}}
\newcommand{\ee}{\end{eqnarray}}
\begin{document}
\date{}
\large
\title{Cosmological and black hole islands in multi-event horizon spacetimes}
\author{${\it Gopal \  Yadav}^{1,}$\footnote{email- gyadav@ph.iitr.ac.in}
~~and~~${Nitin \ Joshi^{2,}}$\footnote{email- 2018phz0014@iitrpr.ac.in}\vspace{0.1in}\\
\textsuperscript{1}{Department of Physics,
Indian Institute of Technology Roorkee, Roorkee 247667, Uttarakhand, India}\\ \textsuperscript{2}{Department of Physics, Indian Institute of Technology Ropar, Rupnagar, Punjab 140001, India}}

\maketitle
\begin{abstract}
In this paper, we have analyzed the information paradox and its resolution using the island proposal in Schwarzschild de-Sitter black hole spacetime. First, we study the information paradox for the black hole patch by treating de-Sitter patch on both sides as a frozen background (by inserting thermal opaque membranes) and then carry out a similar study for de-Sitter patch. In both cases, when there is no island surface,  the entanglement entropy has the linear time dependence as usual, whereas in the presence of an island surface, entanglement entropy becomes constant
(equal to twice of thermal entropy of black hole/de-Sitter patch). Therefore, we obtain the Page curves for the black hole and de-Sitter patches consistent with the unitary evolution of black holes. In our case, we have found that the black hole island is located inside the black hole event horizon in contrast to the universal result for eternal black holes, and the cosmological island is also located inside cosmological event horizon. Further, we have studied the “effect of temperature” on Page curves and found that Page curves appear at late times for low-temperature black hole/de-Sitter patch and exhibit the opposite behavior for high-temperature. This implies that “dominance of islands” and “information recovery” takes more time for low-temperature black hole/de-Sitter patch in contrast to high-temperature black hole/de-Sitter patch. We also comment on the challenges of studying the information paradox in SdS spacetime without the thermal opaque membranes.
\end{abstract}

{\small \bf Keywords :}{\small \vspace{1mm} SdS spacetime, Cosmological and black hole island surfaces, Information paradox, Page curve, Scrambling time}
\newpage	

\tableofcontents

\section{INTRODUCTION}\label{basics}
The phenomenon of black hole evaporation has astonished the world of science for quite some time. This marvelous phenomenon is known as  Hawking radiation ~\cite{Hawking}. There are a lot of papers available in the literature to explore this phenomenon for different spacetimes, e.g., eternal black holes, Rindler or a nonextremal black hole background where the particle creation occurs in causally disconnected spacetime wedges and also for uniformly accelerated or Rindler observers in flat spacetimes known as the Unruh effect \cite{Unruh:1976db} etc. We refer our reader to \cite{DeWitt, Birrell:1982ix, Srinivasan:1998ty, Barman:2017fzh, Traschen:1999zr, Crispino:2007eb, Parker} for an extensive review of these outlooks. We can collect Hawking radiation in terms of entanglement entropy. When a black hole gets hotter, it creates more thermal particles or increases randomness. Entanglement entropy monotonically increases without a bound, but at the end of the evaporation, entanglement entropy must go to zero for an evaporating black hole since the state of a black hole is a pure state. Therefore, this creates the problem of information paradox \cite{stephen}.

We must mention that most problems have been studied for black holes in asymptotically flat spacetimes, but the current observation shows that our universe is going through an accelerated phase. So it is natural to ask how this information paradox problem gets affected by the presence of a  positive cosmological constant $\Lambda$. We are interested in the information paradox problem of a stationary black hole with a positive cosmological constant $\Lambda$ or the Schwarzschild de-Sitter black hole spacetime. Information paradox problem for Schwarzschild de-Sitter black holes is significant because these black holes formed during the early inflationary phase of our universe e.g.,\cite{Bousso:1997wi, Chao:1997em, Bousso:1999ms, Anninos:2010gh}. Keeping in mind the current phase of accelerated expansion of our universe, it also provides an excellent toy model for global structures of isolated black holes of our universe. Also, much like the case of black holes, in de-Sitter space, causally disconnected regions exist. So an observer can only access the parts of the universe bounded by their respective horizon. The Schwarzschild de-Sitter black hole has two event horizons, cosmological event horizon (CEH) and black hole event horizon (BEH). The cosmological event horizon serves as the global causal boundary of the de-Sitter spacetimes. These event horizons possess different temperature thermodynamics compared to $\Lambda \leq 0$ single horizon spacetime ~\cite{Lochan:2018pzs, Goheer:2002vf, Marolf:2010tg}. Like the black hole, cosmological event horizons emit and absorb radiation (Gibbons-Hawking radiation). Loosely speaking, compared to the black hole, the entropy generation of the cosmological horizon is an observer-dependent property. It is generated due to ignorance about what lies beyond the cosmological horizon.

Let us see now how one can resolve the information paradox. We can understand the unitary evolution of the black holes using AdS/CFT correspondence \cite{AdS-CFT}. We can calculate the entanglement entropy of CFT using Ryu-Takayanagi (RT) formula \cite{RT}  and Hubney-Rangamani-Takayanagi (HRT) formula \cite{HRT} for time-independent and time-dependent backgrounds. Authors in \cite{EW} obtained quantum corrections to the Ryu-Takayanagi formula, and the resulting entanglement entropy is known as generalized entropy. We can find quantum extremal surfaces (QES) by extremizing the generalized entropy. One can resolve the black hole information paradox using Island proposal \cite{AMMZ}. The proposal originated from the coupling black hole in JT gravity, including conformal matter (in conformal field theory (CFT)) at the end-of-the-world bane to the same CFT bath in two dimensions. Initially, when there is no island surface, one obtains the linear time dependence of entanglement entropy corresponding to Hawking radiation. At late times, the interior of the black hole becomes part of the entanglement wedge of Hawking radiation. It starts contributing to the entanglement entropy, and island contribution is independent of time. We obtain the Page curve by including these contributions together. Island rule was derived using replica trick from gravitational path integral \cite{rw-1,rw-2}.

For the nonholographic models, we can use s-wave approximation to study the black holes in higher dimensions. In s-wave approximation, we ignore the angular part of the metric and are left with a two-dimensional CFT metric. Therefore we can calculate the entanglement entropy of Hawking radiation from 2D CFT formula given in \cite{CC,CC-1}. There are many papers in literature based on island proposal, e.g., \cite{Island-RNBH,Island-SB,Yu-Ge,Omidi,CLDBH,Tian,Flat-space-black-holes,M-Page,Kim+Nam,Yu et al,Du et al,Chowdhury et al,Yu-2,Seyed et al,I-11,I-12,I-9,I-7,I-6,I-3,I-4,I-2,I-1,I-8,I-new}. For higher derivative theories of gravity, see \cite{NBH-HD,HD-Page Curve-2,RNBH-HD,Ankit}. Island in the context of de-Sitter spacetime has been studied in \cite{Sayantan,I-5,Azarnia,Sybesma,I-dS,Geng-QD,Seo,Geng+Nomura+Sun}. A proposal to resolve the information paradox of an evaporating black hole from the matrix model has been given by authors in \cite{Matrix-Entanglement}. There are many works on information paradox from doubly holographic setups, e.g., \cite{Geng,Geng-2,EPS,GB-2,GB-3} and references therein. Authors studied doubly holographic setup in the context of entanglement negativity in \cite{IITK}\footnote{Thanks to V.~Raj for pointing out their work to us.}. There are works on islands and complexity for multiboundary wormholes with multiple horizons \cite{A1,A2}\footnote{We thank A.~Bhattacharya for pointing out their work to us.}.

The motivation of the paper is to study the information paradox and its resolution in black hole patch and de-Sitter patch by inserting thermal opaque membranes on both sides of the region of interest. We would also like to see the ``effect of temperature'' on the Page curves and scrambling time.

The structure of this paper is organized as follows. In \ref{S2}, we review the basic setup of our theory where we cast the Schwarzschild de-Sitter black hole metric in terms of two Kruskal-like coordinates to extend the spacetime beyond them and give the notion of how to separate black hole event horizon and cosmological event horizon by placing thermal opaque membranes to study both of these horizons separately. We have also discussed the island proposal and bulk gravitational effect near the thermal opaque membrane in \ref{S2}. We split \ref{Island in Schwarzschild de-Sitter Black Hole} into two subsections: in \ref{IP-BH}, we address the information paradox and its resolution using island proposal in black hole patch, and in \ref{IPde-Sitter}, we follow a similar procedure for the cosmological de-sitter patch. In \ref{bhch}, we comment on why it is impossible to study the information paradox problem in the whole SdS spacetime (without the thermal opaque membrane). Finally, in \ref{discussion}, we summarize our work. We shall be working with $(-,+,+,+)$ metric signature in $d = (3+1)$- dimensional spacetime and will set $c$ = $\hslash=k_{\rm B}=1$ throughout.

\section{BASIC SETUP}\label{S2}
\noindent
In the spherical polar coordinates, Schwarzschild de-Sitter(SdS) spacetime metric reads 
\begin{eqnarray}
ds^2=-\left(1-\frac{2M}{r}-\frac{\Lambda r^2}{3}\right)dt^2+\left(1-\frac{2M}{r}-\frac{\Lambda r^2}{3}\right)^{-1}dr^2+r^2 \left(d\theta^2 +\sin^2\theta d\phi^2 \right),
\label{l1}
\end{eqnarray}
where $M$ is the mass parameter. To find the horizons of the above spacetime, we find the points where the timelike Killing vector field becomes null. We need to solve the equation, $\left(1-\frac{2M}{r}-\frac{\Lambda r^2}{3}\right)=0$. It is a cubic equation admitting three roots, for $0<3M \sqrt{\Lambda} < 1$, e.g.~\cite{Gibbons, JHT,SA}.
\begin{eqnarray}
r_{H}=\frac{2}{{\sqrt \Lambda} }\cos\frac{\pi+\cos^{-1}(3M\sqrt{\Lambda})}{3},~
r_{C}=\frac{2}{{\sqrt \Lambda} }\cos\frac{\pi-\cos^{-1}(3M \sqrt{\Lambda})}{3},~r_{U}=-(r_H+r_C),
\label{l2}
\end{eqnarray}
where $r_H, r_C$, are positive, known as the black hole (BEH) and the cosmological event horizon (CEH), respectively, and $r_{U}<0$ is the unphysical horizon. For $3M\sqrt{\Lambda}<1$, \ref{l1} represents a Schwarzschild black hole sitting inside the cosmological horizon. In the Nariai limit, $3M\sqrt{\Lambda} \to 1$, both of the horizons merge $r_H \to r_C$, and for $3M\sqrt{\Lambda}>1$, the spacetime has the naked curvature singularity. The respective surface gravities of the black hole and cosmological event horizons are given as,
\begin{eqnarray}\label{l3}
&& \hskip -0.25 in \kappa_H= \frac{\Lambda (2r_H+r_C)(r_C-r_H)}{6 r_H}=-\sqrt{\Lambda}\left(\cos\left[\frac{1}{3}\cos^{-1}(3M\sqrt{\Lambda})+\frac{\pi}{3}\right]-\frac{1}{4\cos\left[\frac{1}{3}\cos^{-1}(3M\sqrt{\Lambda})+\frac{\pi}{3}\right]}\right),\nonumber\\  
& & \hskip -0.25in -\kappa_C=\frac{\Lambda (2r_C+r_H)(r_H-r_C)}{6 r_C}=\sqrt{\Lambda}\left(\frac{1}{4\cos\left[\frac{1}{3}\cos^{-1}(3M\sqrt{\Lambda})-\frac{\pi}{3}\right]}-\cos\left[\frac{1}{3}\cos^{-1}(3M\sqrt{\Lambda})-\frac{\pi}{3}\right]\right).\nonumber\\
\end{eqnarray}
The negative sign in front of $\kappa_C$ is due to the repulsive effects of positive cosmological constant $\Lambda$. Note that since $r_C\geq r_H$, we have $\kappa_H\geq \kappa_C$, and in the Nariai limit, both the surface gravity $\kappa_H$ and $\kappa_C$ vanish. \ref{l1} contain two coordinate singularities at $r=r_H,\,r_C$, so we need two Kruskal-like coordinates to remove them and extend the spacetime beyond them. First we cast the \ref{l1} into the following form \cite{Bhattacharya:2018ltm}
\begin{eqnarray}
ds^2=\left(1-\frac{2M}{r}-\frac{\Lambda r^2}{3}\right)\left(-dt^2+dr_{\star}^2\right) +r^2(r_{\star})\left(d\theta^2 +\sin^2\theta d\phi^2 \right),
\label{ds4}
\end{eqnarray}
where $r_{\star}$, is the tortoise coordinate, given as
\begin{eqnarray}
r_{\star}=\int \left(1-\frac{2M}{r}-\frac{\Lambda r^2}{3}\right)^{-1}\,dr=\frac{1}{2\kappa_H}\ln \left(\frac{r}{r_H}-1\right) -\frac{1}{2\kappa_C} \ln \left(1-\frac{r}{r_C}\right) +\frac{1}{2\kappa_u}\ln \left(\frac{r}{r_U}-1\right), 
\label{tor}
\end{eqnarray}
where $\kappa_u= \frac{\kappa_H\kappa_C}{\kappa_H-\kappa_C}= (M/r_u^2-\Lambda r_u/3)$ related to the surface gravity of the unphysical horizon $r_u$. Note that if we expand the log's in the above equation, in the limit $\Lambda \to 0$, $r_C\approx \sqrt{\frac{3}{\Lambda}}\to \infty$ or $\kappa_C\approx 1/r_C \to 0$,
we get the Schwarzschild limit, $r_{\star}\approx r+ 2M \ln (r/2M-1)$. Also note that, for $\Lambda \to 0$, we have the expansion for $\kappa_H$ and $r_H$ as,
\begin{eqnarray}
\label{small-KH-rH}
& & 
    \kappa_H = \frac{1}{4 M}-\frac{4 M \Lambda}{3}+ {\cal O}\left(\Lambda\right)^{3/2}\approx \frac{1}{4 M}, \nonumber\\
    & & r_H= 2 M + \frac{8 M^3 \Lambda}{3}+ {\cal O}\left(\Lambda\right)^{3/2}\approx 2 M.
\end{eqnarray}
Now we use $u=t-r_{\star}$, $v=t+r_{\star}$ the retarded and the advanced null coordinates respectively to rewrite~\ref{ds4} as 
\begin{eqnarray}
ds^2=-\left(1-\frac{2M}{r}-\frac{\Lambda r^2}{3}\right)\,dudv +r^2(u,v)\left(d\theta^2 +\sin^2\theta d\phi^2 \right).
\label{ds7}
\end{eqnarray}
Using~\ref{tor}, we have the above metric~\ref{ds7} into two alternative forms,

\begin{eqnarray}
ds^2=-\frac{2M}{r}\left\vert1-\frac{r}{r_C}\right\vert^{1+\frac{\kappa_H}{\kappa_C}} \left(1+\frac{r}{r_H+r_C}\right)^{1-\frac{\kappa_H}{\kappa_U}}\, d{U}_H d {V}_H+r^2(d\theta^2+\sin^2\theta d\phi^2),
\label{ds16}
\end{eqnarray}
and
\begin{eqnarray}
ds^2=-\frac{2M}{r}\left\vert\frac{r}{r_H}-1\right\vert^{1+\frac{\kappa_C}{\kappa_H}} \left(1+\frac{r}{r_H+r_C}\right)^{1+\frac{\kappa_C}{\kappa_U}}\, d {U}_C d {V}_C+r^2(d\theta^2+\sin^2\theta d\phi^2),
\label{ds17}
\end{eqnarray}
where Kruskal null coordinates are define below:
\begin{eqnarray}
{U}_H=-\frac{1}{\kappa_H}e^{-\kappa_H u},\quad {V}_H=\frac{1}{\kappa_H}e^{\kappa_H v} \quad {\rm and} \quad
{U}_C=\frac{1}{\kappa_C}e^{\kappa_C u},\quad {V}_C=-\frac{1}{\kappa_C}e^{-\kappa_C v}.
\label{ds15}
\end{eqnarray}
Now ~\ref{ds16} and ~\ref{ds17} are free of coordinate singularities at  $r=r_H$ and $r_C$. Still, sadly there is no single Kruskal coordinate that simultaneously removes the coordinate singularities of both the horizons for the SdS spacetime. One needs to freeze the other event horizons to study the information paradox in the black hole patch and vice versa. It can be achieved by placing a thermal opaque membrane in between.\par

{\it \textbf{Realization of thermal opaque membranes in SdS spacetime:}} The concept of thermal opaque membrane has been vastly used in literature to study one horizon and take another as the boundary, e.g. \cite{Saida:2009ss,Ma:2016arz,Sekiwa:2006qj,Gomberoff:2003ea} and the references therein. We illustrate the proposal given in \cite{Nitin} of the practical realization of thermal opaque membranes. The Klein-Gordon equation in $3+1$-dimensions, with the radial function, has the following form
\begin{eqnarray}\label{thermalwall}
\left(-\frac{\partial^2}{\partial t^2}+\frac{\partial^2}{\partial r_{\star}^2} \right)R(r)+\left(1-\frac{2M}{r}-\frac{\Lambda r^2}{3} \right)\left(\frac{l(l+1)}{r^2} +\frac{2M}{r^3} -\frac{\Lambda}{3} \right) R(r)=0 .
\end{eqnarray}
An interesting thing to note about the above Schr\"{o}dinger-like equation is that the effective potential term vanishes at both horizons and is positive in between. Thus this bell-shaped potential works as a barricade between the black hole and cosmological event horizons. To visualize it in the Carter-Penrose diagram of the extended Schwarzschild de-Sitter spacetime, we use \ref{ds15} to define the Kruskal timelike and spacelike coordinates as,
$$U_H = T_H -R_H, \quad V_H = T_H+R_H,\qquad {\rm and} \qquad U_C = T_C -R_C, \quad V_C = T_C+R_C, $$
multiplying $U_H$ with $V_H$ and $U_C$ with $V_C$, we have the following relations
\begin{eqnarray}
&&-U_H V_H=R_H^2 -T_H^2= \frac{1}{\kappa_H^2} \left\vert 1-\frac{r}{r_C}\right\vert^{-\kappa_H/\kappa_C} \left\vert \frac{r}{r_U}-1\right\vert^{\kappa_H/\kappa_C}\left(\frac{r}{r_H}-1 \right),\nonumber\\
&&-U_C V_C=R_C^2 -T_C^2= -\frac{1}{\kappa_C^2} \left\vert \frac{r}{r_U}-1\right\vert^{-\kappa_C/\kappa_U} \left\vert \frac{r}{r_H}-1\right\vert^{-\kappa_C/\kappa_H}\left(1-\frac{r}{r_C} \right).
\label{ds5'}
\end{eqnarray}
Thus $r=${\it constant} line is a hyperbola joining $i^{\pm}$, and it can be drawn with respect to any of the above Kruskal coordinates, shown in the Carter-Penrose diagram of the maximally extended SdS spacetime in the following sections. Once we place this hyperbola or the thermal opaque membrane, modes of either side of this wall cannot penetrate it and remain in their respective regions. This effective potential can be a natural realization of the thermally opaque membrane. Now, we shall be discussing the island proposal \cite{AMMZ} to calculate the entanglement entropy of Hawking radiation.

{{\it \textbf {Island rule:}}} One can calculate the entanglement entropy of Hawking radiation using semiclassical generalized entropy formula
\begin{eqnarray}
\label{Island-proposal}
S_{\rm gen}(r)={\rm min}_{\cal I} \Biggl[ {\rm ext}_{\cal I}\Biggl(\frac{Area(\partial {\cal I})}{4 G_N} + S_{\rm matter}({\cal R} \cup {\cal I})\Biggr)\Biggr],
\end{eqnarray}
where ${\cal R}, G_N$, and ${\cal I}$ are the radiation region, Newton constant, and island surface, respectively. As we can see that, generalized entropy consists of two parts: the area of boundary of island surface $ Area (\partial {\cal I})$ and the matter part $S_{\rm matter}({\cal R} \cup {\cal I})$. Suppose there is no island surface, then $S_{\rm gen}(r)=S_{\rm matter}({\cal R})$ (which is entanglement entropy of Hawking radiation outside the black hole). The rule is that first, we need to extremize \ref{Island-proposal} with respect to island coordinates. We must pick the minimal one if there is more than one island surface.\\

\begin{figure}[h!]
\begin{center}
  \includegraphics[width=12.0cm]{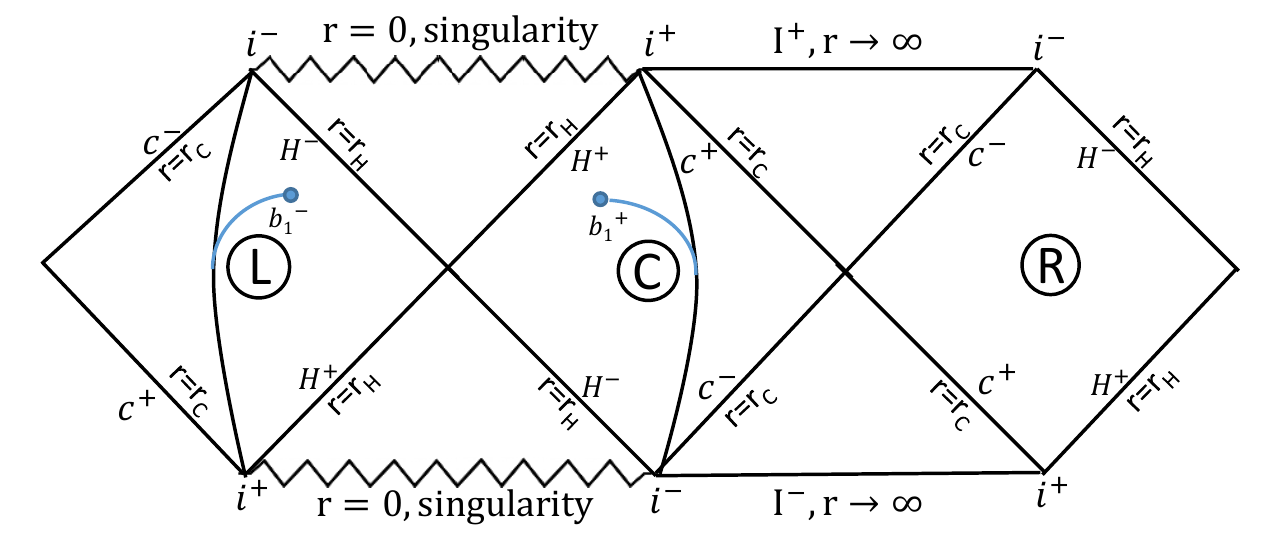}
  \caption{Carter-Penrose diagram of Schwarzschild de-Sitter spacetime. All seven wedges are causally disconnected, and the regions $R, L$ are time reversed with respect to $C$. spacetime can further be extended toward both sides indefinitely for the relevant purpose. Blue color curves are the radiation regions $\mathcal{R}$, introduced on the thermal opaque membranes on both sides. $H^\pm(C^\pm)$ showing the future and past black hole event horizons (cosmological event horizons), $i^\pm$ represent the future and past timelike infinities, respectively and the $I^\pm$ are spacelike infinities.
}
  \label{fig1}
\end{center}
\end{figure}

\begin{figure}[h!]
\begin{center}
  \includegraphics[width=12.0cm]{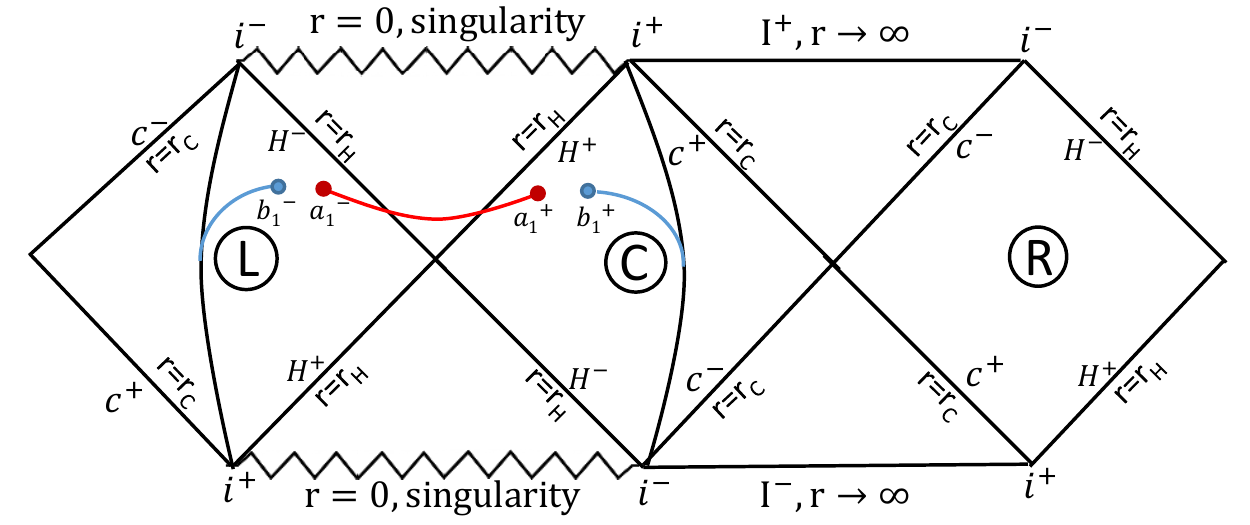}
  \caption{Carter-Penrose diagram of Schwarzschild de-Sitter spacetime. In this figure, we have introduced an island surface $\mathcal{I}$ (red curve) in the black hole patch. Radiation regions $\mathcal{R}$ have been introduced similarly to the previous case on both sides.}
  \label{fig2}
\end{center}
\end{figure}

{\it \textbf {Bulk gravitational effect near thermal opaque membrane:}}
It was discussed in \cite{GB-2,GB-3,GB-4} that one could get the Page curve for a massive theory of gravity in the context of doubly holographic setup because when we consider massless graviton, then island surface does not exist. Hence, there is no Page curve\footnote{See recent work on wedge holography \cite{Miao-MG} where the author has shown that we can get Page curve for massless gravity by adding suitable terms on the end-of-the-world brane, e.g., Dvali-Gabadadze-Porrati (DGP) terms \cite{DGP}. There are other papers too, e.g., \cite{critical-islands,HD-Page Curve-2} where authors were able to obtain the Page curve in massless gravity from doubly holographic setup.}. The calculation of entanglement entropy is based on the factorization of Hilbert space. In quantum field theories, it is an easy job, but Hilbert space does not factorize in gravitational theories \cite{GB-1,GB-5} and one may not get fine-grained Page curve for dynamical gravity \cite{GB-1}. We want to see the bulk gravitational effect on thermal opaque membranes in this discussion. Authors in \cite{Sybesma,anchor-curve} used the island formula to study information paradox for pure de-Sitter space, evaporating and eternal black holes in weak gravity regime. These setups do not contain an external bath; authors have introduced an anchor curve that creates interior and exterior. In the exterior region, gravity is arbitrarily weak enough to use the island formula. Based on \cite{Sybesma,anchor-curve}, we argue that one can think of an ``anchor curve'' passing through the points $b_{1}^{\pm}$ in our setup as well where we have defined the boundaries of radiation regions in  \ref{fig1} and \ref{fig2}, Page curves of Schwarzschild patch is shown in \ref{PC-BH-Lambda}, one
can also draw “anchor curve” for de-Sitter patch too, passing
through points $b_{2}^{\pm}$ in \ref{fig3} and \ref{fig4}. These ``anchor curves'' define the interior and exterior in our setup. The role of the ``thermal opaque membrane'' is to block the radiation from the region, which is not of interest to us. Therefore the validity of the island formula is speculative in our case too. But this is one proposal by which we can resolve the information paradox of black holes in multievent horizon spacetime.

One can also understand the ignorance of gravity near thermal opaque membranes in the language of wedge holography. For this purpose, let us first discuss wedge holography. In wedge holography, we consider two gravitating end-of-the-world branes($B_1$ and $B_2$) of tensions $T_1$ and $T_2$ with $T_1>T_2$. In this setup, $B_1$ contains a black hole that emits Hawking radiation, which is collected to an external gravitating bath $B_2$. The point at which $B_1$ and $B_2$ are joined is known as the ``interface point'' or ``defect''. If end-of-the-world branes are described in $d$ dimensions, then the dimension of ``defect'' will be $d-1$. Wedge holography has the third description: end-of-the-world branes are embedded in $d+1$ dimensions. The key point we want to pick from wedge holography is that gravity is completely absent at the ``defect'' \cite{GB-2}. 

We can think about ``thermal opaque membrane'' in Schwarzschild de-Sitter black holes as the analog of ``defect'' in wedge holography\footnote{We would like to clarify that this is just an analogy because in wedge holography there is transparent boundary condition at interface point and ``thermal opaque membrane'' is nontransparent. The most crucial point is that we have a nonholographic model in our case.}. 
Since the membrane mentioned above is far away from the black hole/de-Sitter patch, therefore, gravity can be assumed to be weak enough near these membranes\footnote{See \cite{Sybesma,anchor-curve} for related works on the application of the Island formula in the weak gravity regime. How to completely switch off gravity on a ``thermal opaque membrane'' is a challenging problem because of the absence of an external bath in this setup.} and 2D CFT formulas can be used to calculate entanglement entropy of Hawking radiation. Although there is no holography in our setup, this is one way of thinking. Based on this point, one trivial question that comes to mind is how one can describe double holography/wedge holography for ``multi-event horizon black holes''. We think this will be an excellent description to resolve the information paradox for black holes in multi-event horizon spacetimes. 
\section{ISLANDS IN SCHWARZSCHILD dE-SITTER BLACK HOLE}\label{Island in Schwarzschild de-Sitter Black Hole}

In this section, we study the information paradox in  Schwarzschild de-Sitter black hole by splitting our discussion into two subsections for black hole patch and cosmological de-Sitter patch. We will use two dimensional CFT formula in s-wave approximation \cite{Island-SB} (observer is very far away from the black hole, i.e, $b_{1,2}^{\pm} \gg r_{H,C}$) to obtain the entanglement entropy of Hawking radiation.

\subsection{\bf{Information paradox in black hole patch}}
\label{IP-BH}

To study the information problem for the black hole patch, we freeze the de-Sitter patch by inserting the aforementioned thermal opaque membrane on both sides. We have defined the boundaries of radiation regions on the thermal opaque membranes on both sides. Metric for the black hole region is\footnote{ We are rewriting metric, conformal factor etc. here, given in \ref{S2}, for the completeness.}
\begin{eqnarray}
\label{metric-BH}
ds^2= - g_H^2(r) dU_H \ dV_H +r^2\left(d\theta^2+\sin^2\theta d\phi^2\right),
\end{eqnarray}
where conformal factor and Kruskal coordinates appearing in \ref{metric-BH} are defined below
\begin{eqnarray}
\label{conf-BH}
& &
g^2_H(r)=\frac{2 M}{r} \biggl|1-\frac{r}{r_C}\biggr|^{\left(1+\frac{\kappa_H}{\kappa_C}\right)}\left(1+\frac{r}{r_C+r_H}\right)^{\left(1-\frac{\kappa_H}{\kappa_U}\right)},\nonumber\\
& & U_H=-\frac{1}{\kappa_H}e^{-\kappa_H \left(t-r_*(r)\right)}; \ V_H=\frac{1}{\kappa_H}e^{\kappa_H \left(t+r_*(r)\right)},
\end{eqnarray}
where tortoise co-ordinate $(r_*(r))$ is given in \ref{tor}. 
Surface gravities for the black hole and de-Sitter horizons are defined in \ref{l3}. The thermal entropy of the black hole turns out to be
\begin{eqnarray}
S_{\rm th}^{\rm BEH}=\frac{A_{\rm BEH}}{4 G_N}=\frac{\pi r_H^2}{G_N}.
\end{eqnarray} 
\subsubsection{\it {\bf No island phase}}
\label{No-Island-BH}
Here, for the black hole patch, our region of interest is the area enclosed by the thermal opaque membrane on both sides in $C$ and $L$ regions.
One can obtain the entanglement entropy of Hawking radiation using the following two-dimensional CFT formula \cite{CC,CC-1} in the absence of the island surface,  
\begin{eqnarray}
\label{EE-formula-no-island-BH}
S_{\rm BEH}^{\mathcal{R}} = \frac{Q}{6} \log\left(d^2(b_1^+,b_1^-)\right),
\end{eqnarray}
where $b_1^+(t_{b_1},b_1)$, $b_1^-(-t_{b_1}+\iota \frac{\beta}{2},b_1)$, (where $\beta=2\pi/\kappa_H$ and $t_{b_1}$ is the boundary time for the radiation region in black hole patch), correspond to the boundaries of radiation regions in the right and left wedges of the black hole patch respectively and $Q$ is the central charge of 2D CFT. Central charge $Q$ take two numeric values, $1$ for bosons and $1/2$ for fermions. Geodesic distance between the two points $l_1$ and $l_2$ for the 2D part of metric \ref{metric-BH} is given by the following expression \cite{NBH-HD}
\begin{equation}
\label{d}
d(l_1,l_2)=\sqrt{g_H(l_1)g_H(l_2)(U_H(l_2)-U_H(l_1))(V_H(l_1)-V_H(l_2))}.
\end{equation}

We can simplify equation \ref{EE-formula-no-island-BH}, using \ref{conf-BH}, \ref{tor} and \ref{d}. The simplified form is written as follows

\begin{eqnarray}
\label{HR-BH-LT}
    S_{\rm BEH}^{\mathcal{R}} = \frac{Q}{6} \log \left(\frac{4 g_H(b_1)^2}{\kappa_H^2} \cosh^2\left(\kappa_H t_{b_1}\right)\right).
\end{eqnarray}
In the late time approximation, i.e., $t_{b_1} \rightarrow \infty$, $\cosh\left(\kappa_H t_{b_1}\right) \sim e^{\kappa_H t_{b_1}}$,
\ref{HR-BH-LT} becomes
\begin{eqnarray}
\label{EE-R-BH}
    S_{\rm BEH}^{\mathcal{R}} \sim \frac{Q}{3}\kappa_H t_{b_1}. 
\end{eqnarray}
The physical significance of the above equation is that due to linear time dependence of entanglement entropy, there will be an infinite amount of Hawking radiation in contrast to the finite amount required by the Page curve of an eternal black hole \cite{Page}.
\subsubsection{\it \textbf{Island phase}}
\label{Island-BH}
In this subsection, we compute the entanglement entropy of Hawking radiation with the inclusion of the island surface in the setup.
The matter part of generalized entropy, \ref{Island-proposal}, which is the entanglement entropy of the Hawking radiation in the presence of the island surface, can be calculated from the following formula\cite{CC,CC-1}
\begin{eqnarray}
\label{EE-formula-island-BH}
S_{\rm matter}^{\rm BEH}({\cal R}\cup {\cal I}) = \frac{Q}{3} \log\left(\frac{d(a_1^+,a_1^-)d(b_1^+,b_1^-)d(a_1^+,b_1^+)d(a_1^-,b_1^-)}{d(a_1^+,b_1^-)d(a_1^-,b_1^+)}\right),
\end{eqnarray}
where $a_1^+(t_{a_1},a_1)$ and $a_1^-(-t_{a_1}+\iota \frac{\beta}{2},a_1)$ are the boundaries of the island surface in the right and left wedges of the black hole patch. In the large distance approximation (observer is very far away from the black hole horizon), $d(a_1^+,a_1^-)\equiv d(b_1^+,b_1^-) \equiv d(a_1^{\pm},b_1^{\mp}) \gg d(a_1^{\pm},b_1^{\pm})$ \cite{Azarnia}, \ref{EE-formula-island-BH} simplifies to
\begin{eqnarray}
\label{EE-formula-island-BH-large-distance}
S_{\rm matter}^{\rm BEH}({\cal R}\cup {\cal I}) = \frac{Q}{6} \log\left(d^2(a_1^+,b_1^+)\right).
\end{eqnarray}
The matter part of generalized entropy (\ref{Island-proposal}) coming from the radiation and island regions can be simplified using \ref{metric-BH}, and \ref{d}, and the simplified expression is
{\footnotesize
\begin{eqnarray}
\label{EE-formula-island-BH-large-distance-simp}
& &
\hskip -0.5in S_{\rm matter}^{\rm BEH}({\cal R}\cup {\cal I}) = \frac{Q}{6} \log\left(\frac{g_H(a_1)g_H(b_1)}{\kappa_H^2}\right)+\frac{Q}{6} \log\left(-\cosh\left(t_{a_1}-t_{b_1}\right) e^{\kappa_H\left(r_*(a_1)+r_*(b_1)\right)}+e^{2 \kappa_H r_*(a_1)}+e^{2 \kappa_H r_*(b_1)}\right).
\end{eqnarray}
}
Hence the generalized entropy (\ref{Island-proposal}), which is a sum of the area of the boundary of the island surface and matter part (\ref{EE-formula-island-BH-large-distance-simp}), is given as
{\footnotesize
\begin{eqnarray}
\label{gen-BH}
& & 
   \hskip -0.5in S_{\rm gen}^{\rm BEH} =\frac{2 \pi a_1^2}{G_N}+\frac{Q}{6} \log\left(\frac{g_H(a_1)g_H(b_1)}{\kappa_H^2}\right)+\frac{Q}{6} \log\left(-2\cosh\left(t_{a_1}-t_{b_1}\right) e^{\kappa_H\left(r_*(a_1)+r_*(b_1)\right)}+e^{2 \kappa_H r_*(a_1)}+e^{2 \kappa_H r_*(b_1)}\right).
\end{eqnarray}
}
Now we are extremizing the generalized entropy for the black hole patch with respect to the island coordinates $(t_{a_1},a_1)$. First we extremize with respect to $t_{a_1}$ in the following way
\begin{eqnarray}
\frac{\partial S_{\rm gen}^{\rm BEH}}{\partial t_{a_1}}=  -\frac{2 e^{\kappa_H\left(r_*(a_1)+r_*(b_1)\right)} \sinh\left(t_{a_1}-t_{b_1}\right)}{e^{2 \kappa_H r_*(a_1)}+e^{2 \kappa_H r_*(b_1) }-2\cosh\left(t_{a_1}-t_{b_1}\right) e^{\kappa_H\left(r_*(a_1)+r_*(b_1)\right)}} =0,
\end{eqnarray}
the solution to the above equation is
\begin{equation}
\label{ta-BH}
t_{a_1} = t_{b_1}+ 2 \iota \pi c_2,
\end{equation}
where $c_2 \in \mathbb{Z}$. Substituting $t_{a_1}$ from above equation into generalized entropy expression \ref{gen-BH}, we obtain
\begin{eqnarray}
    S_{\rm gen}^{\rm BEH} =\frac{2 \pi a_1^2}{G_N}+\frac{Q}{6} \log\left(\frac{g_H(a_1)g_H(b_1)}{\kappa_H^2}\right)+\frac{Q}{6} \log\left(-2 e^{\kappa_H\left(r_*(a_1)+r_*(b_1)\right)}+e^{2 \kappa_H r_*(a_1)}+e^{2 \kappa_H r_*(b_1)}\right).
\end{eqnarray}
The location of the island surface in the black hole patch can be obtained by extremizing the above equation with respect to $a$, i.e.,
\begin{eqnarray}
    \frac{\partial S_{\rm gen}^{\rm BEH}}{\partial a_1}=\frac{4 \pi  r_H}{G_N}+\frac{Q}{4 (a_1- r_H)} =0.
\end{eqnarray}
From the above equation, the location of the island in the black hole patch turns out to be
\begin{eqnarray}
\label{a-BH}
a_1=r_H-\frac{Q G_N}{16 \pi r_H}.
\end{eqnarray}
In this case, the island is located inside the black hole horizon due to the negative sign in front of the second term in the above equation\footnote{Although we have shown the location of islands outside the black hole event horizon in the Penrose diagram, we explicitly found that it is located inside the horizon.}. Now, generalized entropy, after substituting the $a_1$ from \ref{a-BH} turns out to be

\begin{eqnarray}
\label{gen-BH-simp}
    S_{\rm total}^{\rm BEH} =\frac{2 \pi r_H^2}{G_N}+{\cal O}(G_N^0)=2 S_{\rm th}^{\rm BEH}+{\cal O}(G_N^0).
\end{eqnarray}
Therefore we see that when we include the island in the interior of the black hole, the entanglement entropy of Hawking radiation becomes constant, equal to twice the black hole's thermal entropy (up to leading order in $G_N$).\par
{\it \textbf{ Page time:}} The time at which entanglement entropies of Hawking radiation in the absence of the island surface is equal to entanglement entropy in the presence of the island surface is known as Page time. We can get the Page time, by equating \ref{EE-R-BH} and \ref{gen-BH-simp},

\begin{eqnarray}
\label{Page-time-BH}
t_{\rm Page}^{\rm BEH}=\frac{6 S_{\rm th}^{\rm BEH}}{Q \kappa_H}.
\end{eqnarray}
\\
{\it \textbf{Scrambling time:}} The time to recover the information thrown into the black hole in the form of Hawking radiation is known as scrambling time. One can restore the information thrown into the black hole when the black hole has evaporated half of its information \cite{scrambling-time-1,scrambling-time-2}. At late times island comes into the picture, and we have both contributions from the island and the radiation region. This means the entanglement wedge of Hawking radiation consists of radiation and island regions both \cite{scrambling-time-EW}. Hence, in the language of entanglement wedge, the time to reach the information from the boundary of the radiation region to the island surface is scrambling time
\begin{eqnarray}
\label{t-scr}
    t_{\rm scr} = r_*(b_1)-r_*(a_1).
\end{eqnarray}
Therefore for the black hole patch, from \ref{tor} and \ref{a-BH}, scrambling time turns out to be
\begin{eqnarray}
   t_{\rm scr}^{\rm BEH} \approx \frac{1}{2 \kappa_H} \log\left(\frac{\pi r_H^2}{G_N}\right)+ small\approx \frac{1}{2 \kappa_H} \log\left(S_{\rm th}^{\rm BEH}\right) + small.
\end{eqnarray}
%
\subsubsection{\it \textbf{Page curves}}
In this subsection, we use the results of \ref{No-Island-BH} and \ref{Island-BH} to get the Page curves of the black hole patch. The relevant equations are \ref{EE-R-BH} and \ref{gen-BH-simp}. We see that if there is no island surface, then the entanglement entropy of Hawking radiation is proportional to time \ref{EE-R-BH} and when we include the island surface, then it becomes constant $(2 S_{\rm th}^{\rm BEH})$ \ref{gen-BH-simp}. Therefore, when we plot these contributions together, we will obtain the Page curves of eternal black holes consistent with the unitary evolution of the black holes.

\begin{figure}[h!]
\begin{center}
\includegraphics[width=10.0cm]{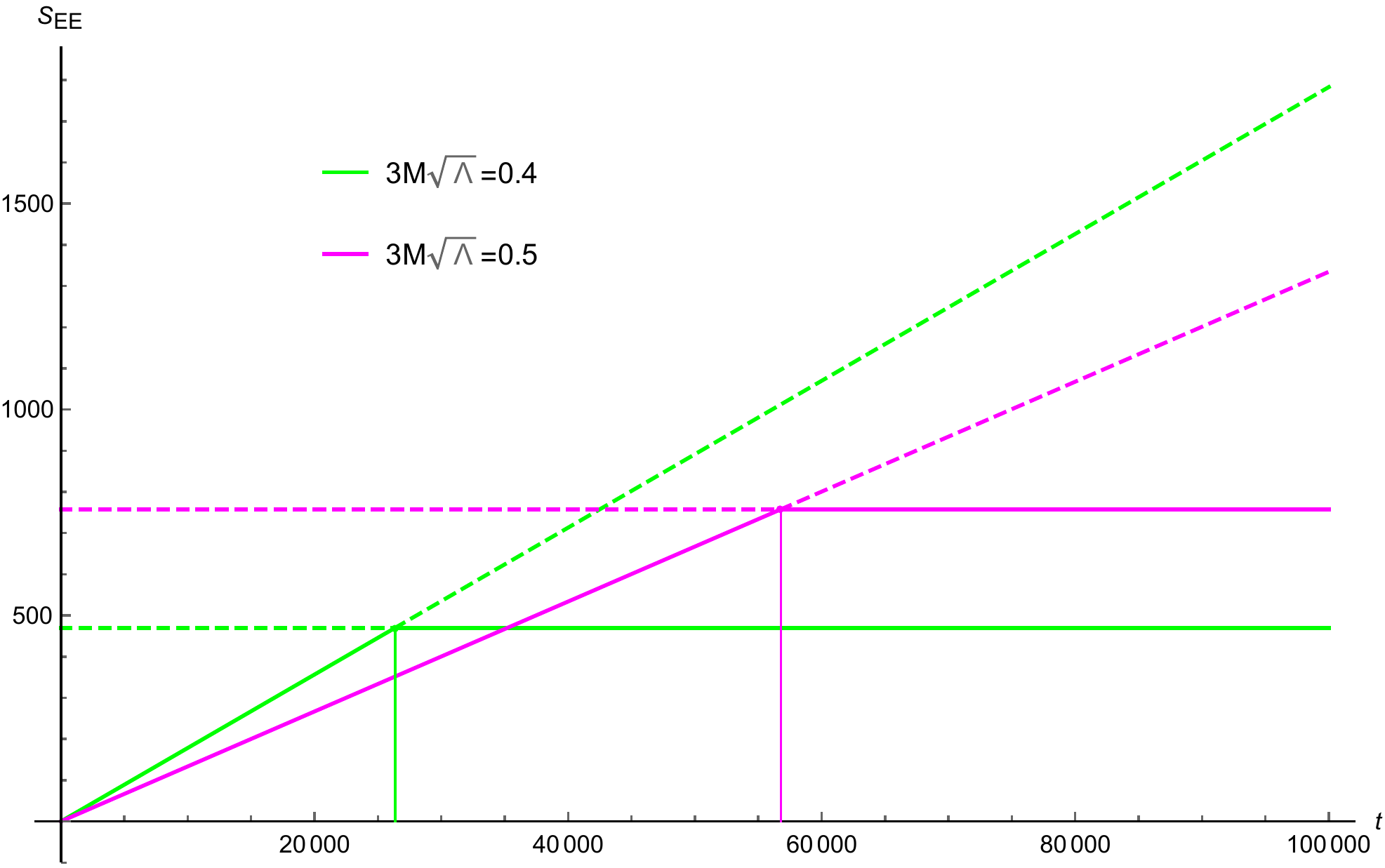}
  \caption{Page curves of black hole patch for fixed values of $3M\sqrt{\Lambda} = 0.4$(green) and $3M\sqrt{\Lambda}=0.5$(magenta). Page times for $3M\sqrt{\Lambda} = 0.4$, $0.5$ black holes are $t_{P_1} \approx 26356.8$  and $t_{P_2} \approx 56787.5$ respectively. }
  \label{PC-BH-Lambda}
\end{center}
\end{figure}

\begin{figure}[h!]
\begin{center}
  \includegraphics[width=12.0cm]{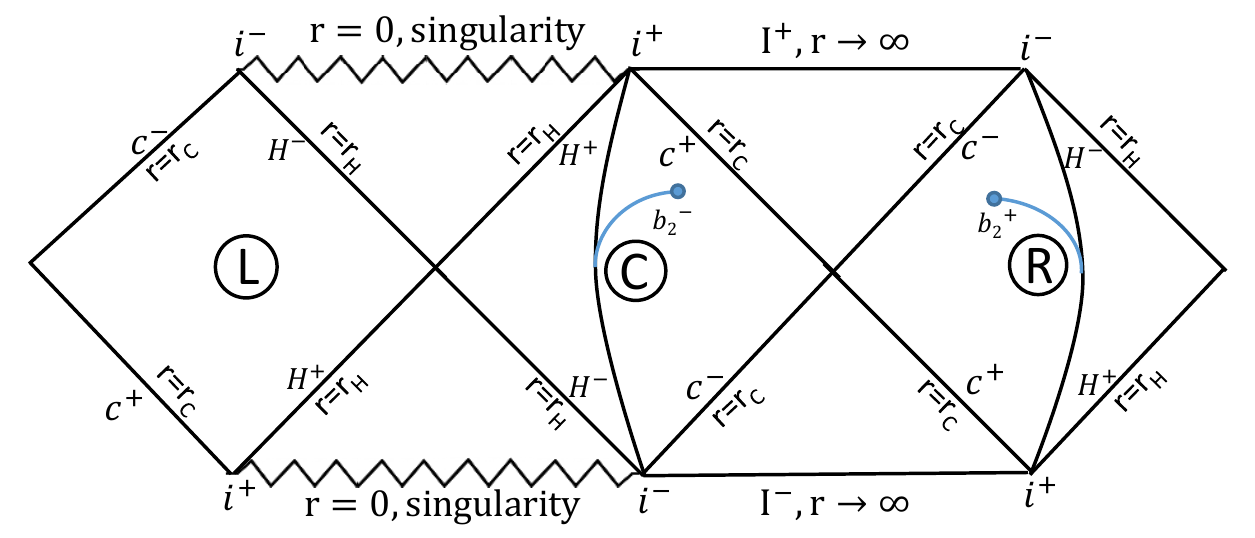}
  \caption{Carter-Penrose diagram of Schwarzschild de-Sitter spacetime. $b^\pm_2$ corresponds to the boundaries of radiation regions $\mathcal{R}$ defined on the thermal opaque membranes similar to the black hole case. See the main text for discussion.}
  \label{fig3}
\end{center}
\end{figure}

\begin{figure}[h!]
\begin{center}
  \includegraphics[width=12.0cm]{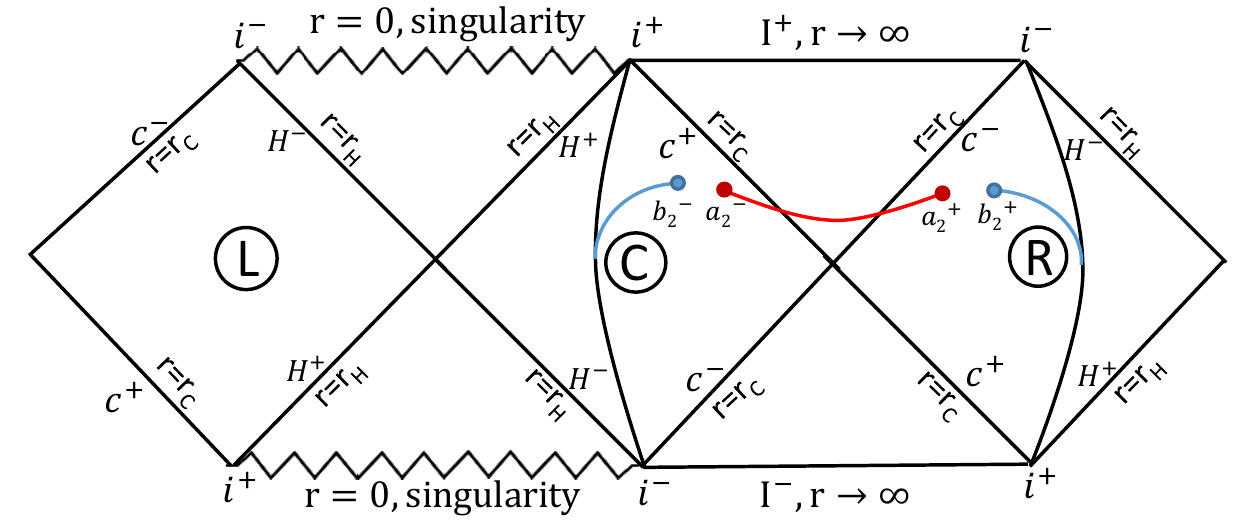}
  \caption{Penrose diagram of Schwarzschild de-Sitter spacetime. In this figure, we have introduced cosmological island surface $\mathcal{I}$ (red curve) in the de-Sitter patch. See the main text for discussion. }
  \label{fig4}
\end{center}
\end{figure}

Now using \ref{l2}, \ref{l3}, \ref{EE-R-BH} and \ref{gen-BH-simp}, we obtain the Page curves of the black hole patch for $3M\sqrt{\Lambda}=0.4$ (green) and $3M\sqrt{\Lambda}=0.5$ (magenta) in \ref{PC-BH-Lambda} (where we set $Q=G_N=1$).

 From \ref{PC-BH-Lambda}, we can see that as the black hole's mass increases, the Page curves are shifting toward later times. Hence, for massive black holes, one must wait a long time to recover the information from the black holes compared to small mass black holes. We can interpret this result in terms of black hole temperature as well. Since black hole temperature is inversely proportional to its mass, we can recover the information earlier in high-temperature black holes compared to low-temperature black holes. In other words, the island dominates earlier in high-temperature black holes compared to the low-temperature black holes.


\subsection{\bf {Information paradox in cosmological de-Sitter patch}}
\label{IPde-Sitter}

In this subsection, we will study the information problem for the de-Sitter patch treating black holes on the both sides as frozen backgrounds (for this purpose again we introduce two thermal opaque membranes on both sides of the black hole patch similar to the previous case). Metric for the region of interest is\footnote { We are rewriting metric, conformal factor etc., here again, given in \ref{S2} for the completeness.} 
\begin{eqnarray}
\label{metric-de-Sitter}
ds^2= - g^2_C(r) U_C \ dV_C +r^2\left(d\theta^2+\sin^2\theta d\phi^2\right),
\end{eqnarray}
where conformal factor and Kruskal coordinates for the de-Sitter patch are defined below
\begin{eqnarray}
\label{de-Sitter-conformal-factor}
& &
g^2_C(r)=\frac{2 M}{r} \biggl|\frac{r}{r_H}-1\biggr|^{\left(1+\frac{\kappa_C}{\kappa_H}\right)}\left(1+\frac{r}{r_C+r_H}\right)^{\left(1+\frac{\kappa_C}{\kappa_U}\right)},\nonumber\\
& & U_C=\frac{1}{\kappa_C}e^{\kappa_C \left(t-r_*(r)\right)}; \ V_C=-\frac{1}{\kappa_C}e^{-\kappa_C \left(t+r_*(r)\right)},
\end{eqnarray}
where $r_*(r)$ is the tortoise coordinate given in \ref{tor}. The thermal entropy of the de-Sitter patch is propositional to its horizon radius, given as
\begin{eqnarray}
S_{\rm th}^{\rm CEH}=\frac{A_{\rm CEH}}{4 G_N} =\frac{\pi r_C^2}{G_N}.
\end{eqnarray}

\subsubsection{\it \textbf{No cosmological island phase}}
\label{No-Island-de-Sitter}

For the de-Sitter patch, our region of interest is the area enclosed by the thermal opaque membranes on both sides in regions $C$ and $R$. When there is no island surface, then matter entanglement entropy appearing in the generalized entropy \ref{Island-proposal} will be given by the formula \cite{CC,CC-1},
\begin{eqnarray}
\label{EE-formula-no-island-de-Sitter}
S_{\rm CEH}^{\mathcal{R}} = \frac{Q}{6} \log[d^2(b_2^+,b_2^-)],
\end{eqnarray}

where $b_2^+(t_{b_2},b_2)$ and $b_2^-(-t_{b_2}+\iota \frac{\beta}{2},b_2)$, ($\beta=2\pi/\kappa_C$ and $t_{b_2}$ is the boundary time for the radiation region in the de-sitter patch), correspond to boundaries of radiation regions in the right and left wedge for the de-Sitter patch. Using \ref{metric-de-Sitter}, \ref{de-Sitter-conformal-factor}, \ref{tor} and \ref{d}\footnote{The formula for geodesic distance between two points for the de-Sitter patch is same as the black hole patch, but with different conformal factor $g_C(r)$ and Kruskal coordinates $U_C(r), V_C(r)$.}, the entanglement entropy in the absence of island surface
\ref{EE-formula-no-island-de-Sitter} simplifies to
\begin{eqnarray}
\label{EE-without-island}
S_{\rm CEH}^{\mathcal{R}} = \frac{Q}{6}\log\Biggl[4 \left(\frac{g_C(b_2)}{\kappa_C}\right)^2 e^{-2 \kappa_C r_*(b_2)} \cosh^2 \kappa_C t_{b_2}\Biggr],
\end{eqnarray}
when $t_b \rightarrow \infty$, i.e., at sufficient late times, we can approximate, $ \cosh \kappa_C t_{b_2} \sim e^{\kappa_C t_{b_2}}$, and hence time-dependent part of \ref{EE-without-island} simplifies to
\begin{eqnarray}
\label{EE-ET-ET-1}
S_{\rm CEH}^{\mathcal{R}} \sim \frac{Q}{3} \kappa_C t_{b_2}.
\end{eqnarray}
The above equation implies that entanglement entropy is proportional to the time when no cosmological island surface exists in the de-Sitter patch. Therefore, at late times, entanglement entropy is divergent. But this should be finite, as suggested in \cite{Page} for eternal black holes. Thereby this corresponds to the information paradox in the de-Sitter patch. Let us now solve it by introducing the cosmological island.

\subsubsection{\it \textbf{Cosmological island phase}}
\label{Island-de-Sitter}

In the presence of cosmological island surface matter part of generalized entropy (\ref{Island-proposal}) is equal to entanglement entropy of two disjoint intervals $(b_2^-,a_2^-)$ and $(a_2^+,b_2^+)$ of 2D CFT \cite{Island-SB}
\begin{eqnarray}
\label{EE-formula-island-de-Sitter}
S_{\rm matter}^{\rm CEH}({\cal R}\cup {\cal I}) = \frac{Q}{3} \log\left(\frac{d(a_2^+,a_2^-)d(b_2^+,b_2^-)d(a_2^+,b_2^+)d(a_2^-,b_2^-)}{d(a_2^+,b_2^-)d(a_2^-,b_2^+)}\right),
\end{eqnarray}
where boundaries of island surface are $a_2^+(t_{a_2},a_2)$ and $a_2^-(-t_{a_2}+\iota \frac{\beta}{2},a_2)$.
If the observer is very far away from the de-Sitter horizon, then \cite{Azarnia}, $d(a_2^+,a_2^-) \equiv d(b_2^+,b_2^-) \equiv d(a_2^{\pm},b_2^{\mp}) \gg d(a_2^{\pm},b_2^{\pm})$, so \ref{EE-formula-island-de-Sitter} simplifies to

\begin{equation}
\label{matter-de-Sitter-EE-simp}
S_{\rm matter}^{\rm CEH}({\cal R}\cup {\cal I}) = \frac{Q}{6} \log\left(d^2(a_2^+,b_2^+)\right).
\end{equation}
From \ref{metric-de-Sitter}, \ref{de-Sitter-conformal-factor}, \ref{tor} and \ref{d}, above equation simplifies to the following form
{\footnotesize
\begin{eqnarray}
\label{matter-simp-de-Sitter}
& &
 \hskip -0.3in S_{\rm matter}^{\rm CEH}({\cal R}\cup {\cal I}) =\frac{Q}{6}\log\Biggl[\frac{g_C(a_2)g_C(b_2)}{\kappa_C^2}\Biggl(-2 e^{-\kappa_C\left(r_*(a_2)+r_*(b_2)\right)} \cosh\left(\kappa_C(t_{a_2}-t_{b_2})\right)+e^{-2 \kappa_C r_*(b_2)}+e^{-2 \kappa_C r_*(a_2)} \Biggr)\Biggr].
\end{eqnarray}
}
Hence generalized entropy in the presence of a cosmological island surface is
\begin{eqnarray}
\label{generalised-entropy-de-Sitter}
    S_{\rm gen}^{\rm CEH}=\frac{2 \pi a_2^2}{G_N}+S_{\rm matter}^{\rm CEH}({\cal R}\cup {\cal I}).
\end{eqnarray}
To get the location of the cosmological island surface, we need to extremize the above equation with respect to the location of island surface $(t_{a_2},a_2)$, i.e.
\begin{eqnarray}
\frac{\partial S_{\rm gen}^{\rm CEH}}{\partial t_{a_2}}=  -\frac{2 e^{-\kappa_C\left(r_*(a_2)+r_*(b_2)\right)} \sinh\left(t_{a_2}-t_{b_2}\right)}{e^{-2 \kappa_C r_*(a_2)}+e^{-2 \kappa_C r_*(b_2) }-2\cosh\left(t_{a_2}-t_{b_2}\right) e^{-\kappa_C\left(r_*(a_2)+r_*(b_2)\right)}} =0,
\end{eqnarray}
the solution to the above equation is
\begin{equation}
\label{ta-de-Sitter}
t_{a_2} = t_{b_2}+ 2 \iota \pi c_1,
\end{equation}
where $c_1 \in \mathbb{Z}$. Using \ref{ta-de-Sitter}, \ref{matter-simp-de-Sitter} takes the following form
\begin{eqnarray}
S_{\rm matter}^{\rm CEH}({\cal R}\cup {\cal I}) =\frac{Q}{6}\log\Biggl[\frac{g_C(a_2)g_C(b_2)}{\kappa_C^2}\Biggr]+\frac{Q}{6}\log\Biggl[\Biggl(-2 e^{-\kappa_C\left(r_*(a_2)+r_*(b_2)\right)} +e^{-2 \kappa_C r_*(b_2)}+e^{-2 \kappa_C r_*(a_2)} \Biggr)\Biggr].
\end{eqnarray}
Therefore, the simplified form of the generalized entropy is
\begin{eqnarray}\label{gen-de-Sitter-simp}
S_{\rm gen}^{\rm CEH}=\frac{2 \pi {a_2}^2}{G_N}+\frac{Q}{6}\log\Biggl[\frac{g_C(a_2)g_C(b_2)}{\kappa_C^2}\Biggr]+\frac{Q}{6}\log\Biggl[\Biggl(-2 e^{-\kappa_C\left(r_*(a_2)+r_*(b_2)\right)} +e^{-2 \kappa_C r_*(b_2)}+e^{-2 \kappa_C r_*(a_2)} \Biggr)\Biggr].
\end{eqnarray}

Now extremization of \ref{gen-de-Sitter-simp} with respect to $a$ will give the location of the cosmological island surface in the de-Sitter patch, i.e.,

\begin{eqnarray}
\frac{\partial S_{\rm gen}^{\rm CEH}}{\partial a_2}=\frac{4 \pi  r_C}{G_N}+\frac{Q}{4 (a_2- r_C)} =0,
\end{eqnarray}
hence the location of the cosmological island surface in the de-Sitter patch turns out to be
\begin{eqnarray}
\label{a-de-Sitter}
a_2=r_C-\frac{Q G_N}{16 \pi r_C}.
\end{eqnarray}
So we can see that the cosmological island is located inside the de-Sitter horizon because ${\cal O}(G_N)$ term is negative in $a_2$. Substituting the $a_2$ from above equation in \ref{gen-de-Sitter-simp}, we obtain the total entanglement entropy, at late times
\begin{eqnarray}
\label{gen-de-Sitter-final}
S_{\rm total}^{\rm CEH}=\frac{2 \pi r_C^2}{G_N}+{\cal O}(G_N^0)=2 S_{\rm th}^{\rm CEH}+{\cal O}(G_N^0).
\end{eqnarray}
The above equation shows that total entanglement entropy is constant in the presence of the cosmological island surface at late times. This constant entropy is twice the thermal entropy of the de-Sitter patch up to the leading order in Newton's constant. \par

{\it \textbf {Page-like time:}} We compute the Page-like time for the de-Sitter patch similar to the black hole patch by equating \ref{EE-ET-ET-1} and \ref{gen-de-Sitter-final}, we get
\begin{eqnarray}
\label{Page-time-de-Sitter}
t_{\rm Page}^{\rm CEH}=\frac{6 S_{\rm th}^{\rm CEH}}{Q \kappa_C}.
\end{eqnarray}
%
{\it \textbf {Scrambling time:}} Scrambling time for the de-Sitter patch can be obtained using equations \ref{tor}, \ref{t-scr} and \ref{a-de-Sitter}
\begin{eqnarray}
    t_{\rm scr}^{\rm CEH} =r_*(b_2)-r_*(a_2) \approx \frac{ \log\left(\frac{G_N}{\pi r_C^2}\right)}{ 2 \kappa_C}+small \approx \frac{1}{2 \kappa_C} \log\left(S_{\rm th}^{\rm CEH}\right) + small.
\end{eqnarray}
The above equation implies that de-Sitter horizons act like fastest scramblers similar to the black holes \cite{scrambling-time-2}.

\subsubsection{\it \textbf{Page-like curves}}

In \ref{No-Island-de-Sitter} and \ref{Island-de-Sitter}, we observed that when there is no cosmological island surface, then entanglement entropy of de-Sitter radiation is time-dependent (\ref{EE-ET-ET-1}). When we include the cosmological island surface inside the de-Sitter horizon, entanglement entropy turns out to be constant (\ref{gen-de-Sitter-final}). We can plot these entanglement entropies together to obtain the Page-like curves for the de-Sitter patch. We have used the numerical values, $Q=G_N=1$, cosmological constant $\Lambda =0.001$ and plotted two Page curves for $3 M\sqrt{\Lambda} = 0.4, 0.7$. Page time for $3 M\sqrt{\Lambda} = 0.4$ is $t_{P_1}=3.143 \times 10^6$ and for $3 M\sqrt{\Lambda} = 0.7$ is  $t_{P_2}=3.360 \times 10^6$.
\begin{figure}[h!]
\begin{center}
  \includegraphics[width=10.0cm]{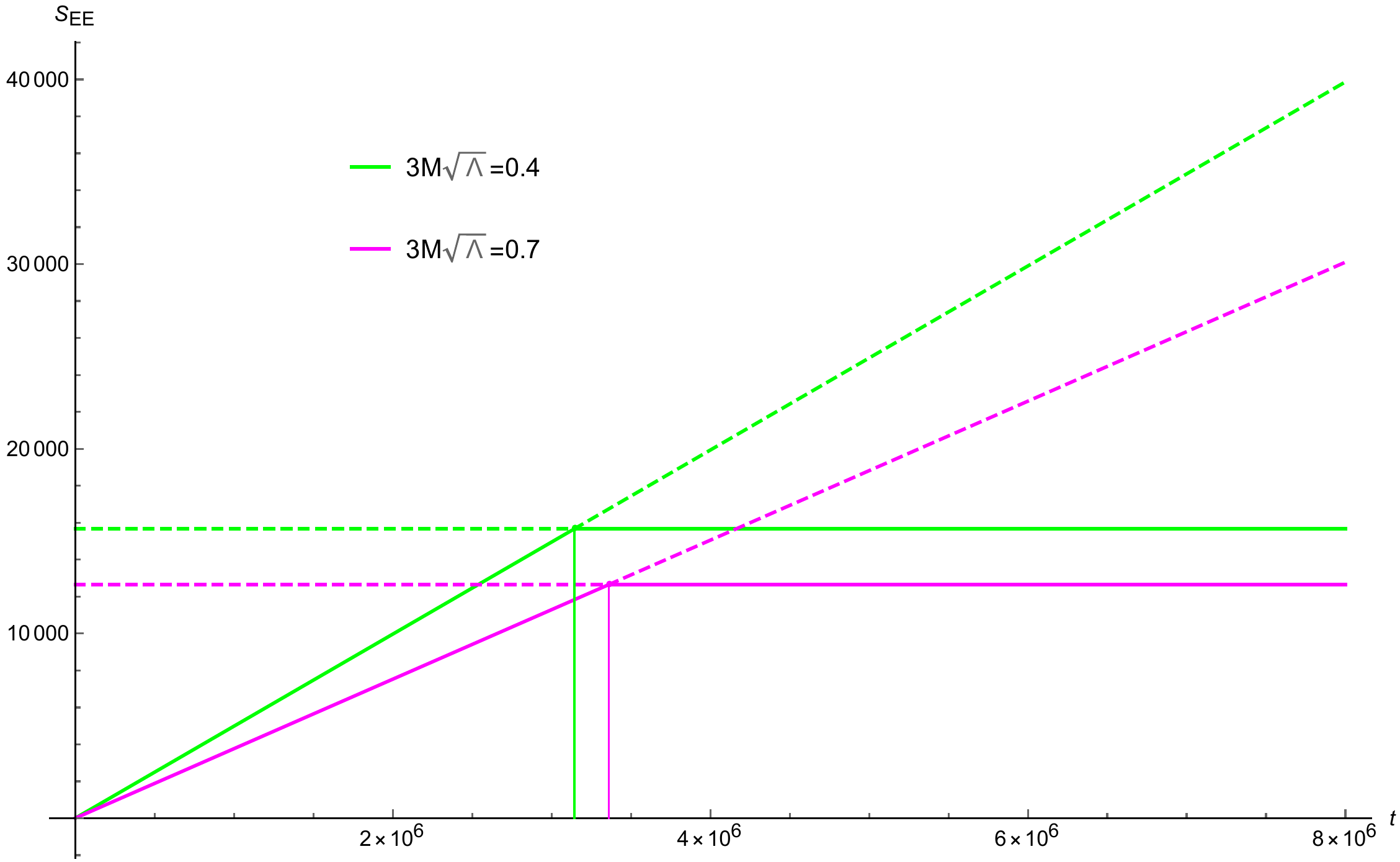}
  \caption{Page curves of de-Sitter patch for $3 M\sqrt{\Lambda} = 0.4$ (green) and $0.7$ (magenta). Page times for $3 M\sqrt{\Lambda} = 0.4$ and $0.7$ are $t_{P_1}=3.143 \times 10^6$ and  $t_{P_2}=3.360 \times 10^6$ respectively. See the main text for discussion.}
  \label{PC-de-Sitter}
\end{center}
\end{figure}

\ref{PC-de-Sitter} represents the effect of a  parameter, $3 M\sqrt{\Lambda}$, on the Page curves of de-Sitter patch. Diagonal green and magenta lines are corresponding entanglement entropies without cosmological island surface (\ref{EE-ET-ET-1}). Constant green and magenta lines correspond to entanglement entropies with cosmological island surface (\ref{gen-de-Sitter-final}) (we have restricted ourselves to leading order in $G_N$) for $3 M\sqrt{\Lambda} = 0.4$ and  $0.7$. Effect of the aforementioned parameter on the Page curves of the de-Sitter patch, if we increase $3M\sqrt{\Lambda}$ Page curves shift toward later times and vice versa. In terms of temperature, this implies that the de-Sitter patch with lower temperature takes a long time to emit the radiation. Hence, we shall be able to recover the information thrown into the de-Sitter patch after a long time for a lower temperature de-Sitter patch than the higher temperature de-Sitter patch.

\section{INFORMATION PARADOX IN MULTI-EVENT HORIZONS SPACETIME}\label{bhch}

One can also think of information paradox problem in the entire Schwarzschild de-Sitter (SdS) spacetime since it possesses an effective equilibrium temperature \cite{Saida:2009ss,Nitin,pc,saida,urano,pappas}. Let us see now what the challenges are here. We have already removed the coordinate singularity at $r=r_U$ and $r=r_C$ in \ref{S2}. Following a similar process, we use \ref{tor} to remove the ``singularity'' of the metric \ref{l1} at $r=r_U$. The $t-r$ part of \ref{l1} given as 
\begin{eqnarray}
ds^2=-\frac{2M}{r}\left\vert1-\frac{r}{r_C}\right\vert^{1+\frac{\kappa_U}{\kappa_C}} \left\vert\frac{r}{r_H}-1\right\vert^{1-\frac{\kappa_U}{\kappa_H}}\, dU d V,
\label{c4}
\end{eqnarray}
where we have,
\begin{eqnarray}
U=-\frac{1}{\kappa_U}e^{-\kappa_U u},\quad V=\frac{1}{\kappa_U}e^{\kappa_U v} \quad .
\label{c5}
\end{eqnarray}

 The above metric is not regular on or across the black hole and cosmological event horizons. Hence it is only confined to the region $C$. So there is no way to write a Kruskal-like coordinates system that is regular on the black hole, cosmological event horizons, and covers the entire SdS spacetime.

 Furthermore, in terms of island proposal, one may discuss the information paradox for the whole SdS spacetime by locating the observers in regions L and R to collect the Hawking radiation, and one can define the boundaries, $(b_-,b_+)$, of radiation regions $({\cal R})$ there. Then one may include two islands, ${\cal I}_1$ and ${\cal I}_2$, with boundaries, $(a_1^-,a_1^+)$ and $(a_2^-,a_2^+)$, in black hole and de-Sitter patches simultaneously. If one follows this procedure, then the matter contribution to the entanglement entropy can be calculated using 2D CFT formula for three intervals as given below\footnote{Without island, matter part will be similar to \ref{IP-BH} and \ref{IPde-Sitter}.}
{\footnotesize
\begin{eqnarray}
\label{EE-3I}
 S_{\rm matter}({\cal R}\cup {{\cal I}_1}\cup {\cal I}_2)=\frac{Q}{3} \log \Biggl(\frac{d(b_-,a_1^-)d(a_1^+,a_1^-)d(a_2^+,a_1^-)d(b_-,a_2^-)d(a_1^+,a_2^-)d(a_2^+,a_2^-)d(b_-,b_+)d(a_1^+,b_+)d(a_2^+,b_+)}{d(a_1^+,b_-)d(a_2^+,b_-)d(a_2^+,a_1^+)d(a_2^-,a_1^-)d(b_+,a_1^-)d(b_+,a_2^-)}\Biggr).
\end{eqnarray}
}
In this case, the first term in generalized entropy (\ref{Island-proposal}) will be of the form $\frac{2 \pi \left(a_1^2+a_2^2\right)}{G_N}$. One can follow a similar procedure to obtain the Page curves as done in \ref{IP-BH} and \ref{IPde-Sitter}. The issue is that if we consider the entire SdS spacetime, then we may not be able to define the same radiation regions on both sides similar to \ref{IP-BH} and \ref{IPde-Sitter} because, on one side, we have black hole patch. In contrast, on the other side, we have the de-Sitter patch. Let us now summarize our work in the next section.

\section{CONCLUSION}\label{discussion}
In this paper, we have studied the information paradox in the context of Schwarzschild de-Sitter black hole spacetime using the Island proposal. We have introduced thermal opaque membranes to isolate the black hole from the de-Sitter patch on both sides and vice-versa. We have computed the entanglement entropy of Hawking radiation in the absence and presence of an island surface and found that, as usual, entanglement entropy has linear time dependence when there is no island surface, and it becomes constant in the presence of an island surface. Therefore, we obtain the Page curve consistent with the unitary evolution of the black holes. In our case, the island is located inside the black hole horizon compared to the universal result that the island is located outside the horizon for eternal black holes \cite{island-o-h}. However, for the eternal black holes, the aforementioned problem can be solved by the quantum focusing conjecture (QFC) \cite{QFC}. Due to QFC, when we decouple or couple a black hole to a bath, then a finite amount of produced energy flux shifts the horizon outward without affecting the island surface's location. Hence, the island always lies inside the black hole horizon. Our result is also similar to \cite{I-3}, where authors computed the Page curve of a one-sided asymptotically flat black hole and discovered that the island is located inside the horizon.
 \par
 
As discussed earlier, radiation emitted by the cosmological event horizon is analogous to Hawking radiation of a black hole. We used the 2D CFT formula \cite{CC,CC-1} even for the de-Sitter patch in s-wave approximation \cite{Island-SB}. Boundaries of the radiation regions for the de-Sitter patch are defined similarly to the black hole case. The de-Sitter patch entanglement entropy also has a linear time dependence when there is no cosmological island surface implying there will be an infinite amount of de-Sitter radiation at late times. Consequently, we will have de-Sitter information paradox (similar to the information paradox of the black hole). To resolve this paradox, we have introduced a cosmological island surface in the de-Sitter patch. The inclusion of the cosmological island surface leads to saturation of the entanglement entropy of de-Sitter radiation, and this saturation value is equal to twice of thermal entropy of de-Sitter radiation. The dominance of entanglement entropy of de-Sitter radiation is similar to Hawking radiation, i.e., before the Page time, entanglement entropy without cosmological island surface dominates. After the Page time, entanglement entropy with the inclusion of cosmological island surface dominates. Hence, we obtain the Page-like curves for the de-Sitter patch as well. Authors in \cite{SdS-BH-Island} studied the information paradox and its resolution for the black hole by locating the observers at the cosmological horizons on both sides. In contrast, in our case, instead of locating the observers at the de-Sitter horizons, here we locate them at thermal opaque membranes.\par

It is interesting to note that even for the de-Sitter patch, the location of the island surface is inside the cosmological event horizon. We have also computed the scrambling times for the black hole and the de-Sitter patch. For both cases, scrambling time is proportional to the logarithmic of their thermal entropy. Therefore de-Sitter patch also works as the fastest scrambler, similar to the black hole \cite{scrambling-time-2}. While the information paradox in the whole SdS spacetime could not be possible since there is no Kruskal-like coordinates system that is regular on both the horizons and cover the entire Schwarzschild de-Sitter black hole spacetime. Also, it is unclear how to define the boundaries of radiation regions for the entire SdS spacetime.

Further, we have analyzed the ``effect of temperature'' on the Page curves of the black hole and de-Sitter patches separately. We found that as the temperature of the black hole/de-Sitter patch increases, Page curves shift toward later times\footnote{Similar results, i.e.,  {\it ``appearance of Page curves at later times or earlier times''}, was also observed in \cite{RNBH-HD,Ankit} due to the presence of higher derivative terms in the gravitational action.}. Therefore black hole/de-Sitter patches with a lower temperature will take a long time to deliver the information to the observer compared to the higher temperature. In terms of the dominance of the island, this implies that the black hole/de-Sitter patch island appears late for the lower temperature black hole/de-Sitter patch, and the island appears earlier for the higher temperature. As soon as an island surface comes into the picture, we start recovering the information thrown into the black hole and cosmological event horizons \cite{scrambling-time-1,scrambling-time-2}. Hence, we conclude that the ``dominance of islands'' and scrambling time depend upon the temperature of both event horizons.

The future direction of this paper could be the study of the complexity of Schwarzschild de-Sitter black holes. It will help us to understand the black hole's interior in detail. We shall be returning to this issue in our future publications.

\section*{ACKNOWLEDGEMENTS}
G.Y. is supported by a Senior Research Fellowship (SRF) from the Council of Scientific and Industrial Research (CSIR), Govt. of India. We would like to thank the organizers of PASCOS 2022 for organizing such a wonderful conference from where we started our collaboration. N.J. would also like to thank Sourav Bhattacharya for his fruitful comments.

\end{document}